\documentclass{appolb}
\usepackage{epsfig}

\usepackage{graphicx}
\usepackage{amsmath}
\usepackage{array}

\def\lg{{\mathchoice{~\raise.58ex\hbox{$<$}\mkern-14.8mu\lower.52ex\hbox{$>$}~}
                    {~\raise.58ex\hbox{$<$}\mkern-14.8mu\lower.52ex\hbox{$>$}~}
                    {\raise.59ex\hbox{{$\scriptscriptstyle <$}}\mkern-12.8mu%
                     \lower.01ex\hbox{{$\scriptscriptstyle >$}}}   {}   }}
\def\gl{{\mathchoice{~\raise.58ex\hbox{$>$}\mkern-12.8mu\lower.52ex\hbox{$<$}~}
                    {~\raise.58ex\hbox{$>$}\mkern-12.8mu\lower.52ex\hbox{$<$}~}
                    {\raise.62ex\hbox{{$\scriptscriptstyle >$}}\mkern-12.0mu%
                     \lower.05ex\hbox{{$\scriptscriptstyle <$}}}  {}    }}

\newcommand{\be}{\begin{equation}}
\newcommand{\ee}{\end{equation}}
\newcommand{\ba}{\begin{eqnarray}}
\newcommand{\ea}{\end{eqnarray}}
\newcommand{\ban}{\begin{eqnarray*}}
\newcommand{\ean}{\end{eqnarray*}}

\newcommand{\sla}{\!\!\!/ \,}

\begin{document}
% \eqsec  % uncomment this line to get equations numbered by (sec.num)

\title{Universal Hard-Loop Actions\thanks{Presented by A. Czajka at 
{\it Excited QCD 2015}, Tatranska Lomnica, Slovakia, March 8-14, 2015.}}

\author{Alina Czajka
\address{Institute of Physics, Jan Kochanowski University, Kielce, Poland}
\and
Stanis\l aw Mr\' owczy\' nski
\address{Institute of Physics, Jan Kochanowski University, Kielce, Poland \\
and National Centre for Nuclear Research, Warsaw, Poland}
}

\maketitle
\begin{abstract}
The effective actions of gauge bosons, fermions and scalars, which are obtained within the hard-loop approximation, are shown to have unique forms for a whole class of gauge theories including QED, scalar QED, super QED, pure Yang-Mills, QCD, super Yang-Mills. The universality occurs irrespective of a field content of each theory and of variety of specific interactions. Consequently, the long-wavelength or semiclassical features of plasma systems governed by these theories such as collective excitations are almost identical. An origin of the universality, which holds within the limits of applicability of the hard-loop approach, is discussed.

\end{abstract}


% Relativistic plasmas, 52.27.Ny
% Supersymmetry, 11.30.Pb
% Quantized fields, 03.70.+k
  
%%%%%%%%%%%%%%%%%%%%%%%%%%%%%%%%%%%%%%%%%%%%%
\section{Introduction}
%%%%%%%%%%%%%%%%%%%%%%%%%%%%%%%%%%%%%%%%%%%%%

The hard-loop approach is a practical tool to describe plasma systems governed by QED or QCD in a gauge invariant way which is free of infrared divergences, see the reviews \cite{Thoma:1995ju,Blaizot:2001nr,Litim:2001db,Kraemmer:2003gd}. Initially the approach was developed within the thermal field theory \cite{Braaten:1989mz,Braaten:1990az} but it was soon realized that it can be formulated in terms of quasiclassical kinetic theory \cite{Blaizot:1993be,Kelly:1994dh}. The plasma systems under consideration were assumed to be in thermodynamical equilibrium but the methods can be naturally generalized to plasmas out of equilibrium \cite{Pisarski:1997cp,Mrowczynski:2000ed}. 

An elegant and concise formulation of the hard-loop approach is achieved by introducing an effective action derived for equilibrium and non-equilibrium systems in \cite{Taylor:1990ia,Frenkel:1991ts,Braaten:1991gm} and \cite{Pisarski:1997cp,Mrowczynski:2004kv}, respectively. The action is a key quantity that encodes an infinite set of hard-loop $n$-point functions. A whole gamut of long-wavelength characteristics of a plasma system is carried by the functions. In particular, the two-point functions or self-energies provide response functions like permeabilities or susceptibilities which control various screening lengths. The self-energies also determine a spectrum of collective excitations (quasiparticles) that is a fundamental characteristic of any many-body system. 

One wonders how much a given plasma characteristic is different for different plasma systems. It has been known for a long time that the self-energies of gauge bosons in the long-wavelength limit are of the same structure for QED and QCD plasmas \cite{Weldon:1982aq}. Consequently, the collective excitations and many other characteristics are the same, or almost the same, in the two plasma systems \cite{Mrowczynski:2007hb}. We show here that the similarity of long-wavelength characteristics is not limited to QED and QCD but there is a whole class of gauge theories which have the universal hard-loop actions of gauge bosons, fermions, and scalars. The class includes: QED, scalar QED, $\mathcal{N}=1$ super QED, pure Yang-Mills, QCD, and $\mathcal{N}=4$ super Yang-Mills. However, the universality occurs only in the domain of validity of the hard-loop approach that is when the momentum scale of collective degrees of freedom is neither too long nor too short. 

This paper is based on our study \cite{Czajka:2014gaa} which provides a comprehensive discussion on different aspects of the universality. Further computational details can be found also in \cite{Czajka:2010zh,Czajka:2011zn,Czajka:2012gq}, where supersymmetric plasmas were systematically compared to their non-supersymmetric counterparts. Throughout the paper we use the natural system of units with $c= \hbar = k_B =1$; our choice of the signature of the metric tensor is $(+ - - -)$.

%%%%%%%%%%%%%%%%%%%%%%%%%%%%%%%%%%%%%%%%%%%%%
\section{Self-energies}
\label{sec-self-en}
%%%%%%%%%%%%%%%%%%%%%%%%%%%%%%%%%%%%%%%%%%%%%

Our aim is to derive the effective action of all considered theories in the hard-loop approximation. The action $S$ can be found {\it via} the respective self-energies which are the second functional derivatives of $S$ with respect to the given fields. Thus, the self-energies of gauge boson $(A^\mu)$, fermion $(\Psi)$ and scalar $(\Phi)$ fields equal
\ba
\label{se-Pi}
\Pi^{\mu \nu}(x,y) &=& \frac{\delta^2 S}{\delta A_\mu(x) \,\delta A_\nu(y)}, 
\\ [2mm]
\label{se-Sigma}
\Sigma (x,y) &=& \frac{\delta^2 S}{\delta \bar\Psi (x) \,\delta \Psi(y)}, 
\ea
\ba
\label{se-P}
P(x,y) &=& \frac{\delta^2 S}{\delta \Phi^*(x) \,\delta \Phi (y)},
\ea
where the field indices, which are different for different theories under consideration, are suppressed. The action will be obtained in the subsequent section by integrating the formulas (\ref{se-Pi})-(\ref{se-P}) over the respective fields. 

The self-energies entering Eqs.~(\ref{se-Pi})-(\ref{se-P}) have been found diagrammatically, for details see \cite{Czajka:2010zh,Czajka:2011zn,Czajka:2012gq}. The plasma systems under study have been assumed to be homogeneous in coordinate space (translationally invariant), locally colorless and unpolarized, but the momentum distribution may be arbitrary. Therefore, we used the Keldysh-Schwinger or real-time formalism, explained in {\it e.g.} \cite{Mrowczynski:1992hq}, which allows one to describe many-body systems both in and out of equilibrium. 

Elaborating on self-energies of a given field of all plasma systems under study one can observe that both the number of diagrams contributing to self-energies and their forms are very different for each theory. Accordingly, there is no surprise that, say, the polarization tensor $\Pi^{\mu \nu}(k)$ is quite different for each theory. However, when the external momentum $k$ is much smaller than the internal momentum $p$, which flows along the loop and is carried by a plasma constituent, that is when the hard-loop approximation $(k \ll p)$ is applied, we get a very striking result: the (retarded) self-energies of gauge boson, fermion, and scalar fields of all theories are of the same form 
\be
\label{Pi-k-final}
\Pi^{\mu \nu}(k)
= C_{\Pi}
\int \frac{d^3p}{(2\pi)^3}
\frac{f_{\Pi}({\bf p})}{E_p} 
\frac{k^2 p^\mu p^\nu - (k^\mu p^\nu + p^\mu k^\nu - g^{\mu \nu} (k\cdot p))
(k\cdot p)}{(k\cdot p + i 0^+)^2},
\ee
\ba
\label{Si-k-final}
\Sigma(k) &=& C_{\Sigma}
\int \frac{d^3p}{(2\pi )^3}
\frac{f_{\Sigma}({\bf p})}{E_p}  \, \frac{p\sla}{k\cdot p + i 0^+},
\ea
\be
\label{P-k-final}
P(k) = - C_P \int \frac{d^3p}{(2\pi)^3} \frac{f_P({\bf p})}{E_p},
\ee
where $C_{\Pi}$, $C_{\Sigma}$, and $C_P$ are some numerical factors and $f_{\Pi}({\bf p})$, $f_{\Sigma}({\bf p})$, and  $f_P({\bf p})$ are the effective distribution functions of plasma constituents of a given system. Both the factors and distribution functions are presented in detail for each plasma system in paper \cite{Czajka:2014gaa} .

The universal expressions of the self-energies (\ref{Pi-k-final}), (\ref{Si-k-final}), and (\ref{P-k-final}) have been obtained in the hard-loop approximation when the external momentum $k$ is much smaller than the internal momentum $p$. However, it appears that the self-energies (\ref{Pi-k-final}), (\ref{Si-k-final}), and (\ref{P-k-final}) are valid when the external momentum $k$ is not too small. It is most easily seen in case of the fermion self-energy (\ref{Si-k-final}) which diverges as $k \rightarrow 0$. When we deal with an equilibrium (isotropic) plasma of the temperature $T$, the characteristic momentum of (massless) plasma constituents is of the order $T$. One observes that if the external momentum $k$ is of the order $g^2 T$, which is the so-called {\it magnetic} or  {\it ultrasoft} scale, the self-energy (\ref{Si-k-final}) is not perturbatively small as it is of the order ${\cal O}(g^0)$. Therefore, the expression (\ref{Si-k-final}) is meaningless for $k \le g^2 T$.  Since $k$ must be much smaller than $p \sim T$, one arrives to the well-known conclusion that the self-energy (\ref{Si-k-final}) is valid at the {\it soft} scale that is when $k$ is of the order $gT$.  Analyzing higher order corrections to the self-energies (\ref{Pi-k-final}), (\ref{Si-k-final}), (\ref{P-k-final}), one shows that they are indeed valid for $k \sim gT$ and they break down at the magnetic scale because of the infrared problem of gauge theories, see {\it e.g.} \cite{Lebedev:1989ev} or the review \cite{Kraemmer:2003gd}. When the momentum distribution of plasma particles is anisotropic, instead of the temperature $T$, we have a characteristic four-momentum ${\cal P}^\mu$ of plasma constituents and the hard-loop approximation requires that ${\cal P}^\mu \gg k^\mu$ which should be understood as a set of four conditions for each component of the four-momentum $k^\mu$. Validity of the self-energies (\ref{Pi-k-final}), (\ref{Si-k-final}), and (\ref{P-k-final}) is then limited to $k^\mu \sim g {\cal P}^\mu$.  

%%%%%%%%%%%%%%%%%%%%%%%%%%%%%%%%%%%%%%%%%%%%%
\section{Effective action}
\label{sec-eff-act}
%%%%%%%%%%%%%%%%%%%%%%%%%%%%%%%%%%%%%%%%%%%%

Having the self-energies $\Pi^{\mu\nu}(k),\; \Sigma(k)$, and $P(k)$ given by Eqs.~(\ref{Pi-k-final}), (\ref{Si-k-final}), and (\ref{P-k-final}), respectively, we can reconstruct the effective action. The effective actions are obtained by integrating the formulas (\ref{se-Pi})-(\ref{se-P}) over the respective fields and next inserting the respective self-energies. Taking into account some gauge symmetry arguments we get 
\ba
\label{action-A-HL}
{\cal L}^{A}_{\rm HL}(x) &=& C_\Pi \int \frac{d^3p}{(2\pi )^3} \,
\frac{f_\Pi({\bf p})}{E_p} \,
F_{\mu \nu} (x) {p^\nu p^\rho \over (p \cdot D)^2} F_\rho^{\mu} (x) ,
\\ [2mm]
\label{action-Psi-HL}
{\cal L}^{\Psi}_{\rm HL}(x) &=& C_\Sigma
\int \frac{d^3p}{(2\pi )^3} \, \frac{ f_\Sigma({\bf p})}{E_p} \,
\bar{\Psi}(x) {p \cdot \gamma \over p\cdot D} \Psi(x) ,
\\ [2mm]
\label{action-Phi-HL}
{\cal L}^{\Phi}_{\rm HL}(x) &=& - C_P
\int \frac{d^3p}{(2\pi )^3} \, \frac{f_P({\bf p})}{E_p} \;
\Phi^*(x) \Phi(x) .
\ea
where the field indices are omitted to keep the expressions applicable to all considered theories. The action is obviously related to the Lagrangian density as $S = \int d^4x \, {\cal L}$. The forms of covariant derivatives present in Eqs.~(\ref{action-A-HL}) and (\ref{action-Psi-HL}) depend on the theory under consideration. In the electromagnetic theories, the derivative in the gauge boson action (\ref{action-A-HL}) is the usual derivative while that in the fermion action  (\ref{action-Psi-HL}) is $D^\mu = \partial^\mu - ie A^\mu$. In the ${\cal N}=4$ super Yang-Mills the covariant derivatives in Eqs.~(\ref{action-A-HL}) and (\ref{action-Psi-HL})  are both in the adjoint representation of ${\rm SU}(N_c)$ gauge group. In QCD, the covariant derivative in Eq.~(\ref{action-A-HL}) is in the adjoint representation but that in Eq.~(\ref{action-Psi-HL}) is in the fundamental one. In case of ${\cal N}=4$ super Yang-Mills there is an extra factor 1/2 in the r.h.s. of Eq.~(\ref{action-Phi-HL}).

The hard-loop actions (\ref{action-A-HL}), (\ref{action-Psi-HL}), and (\ref{action-Phi-HL}) are all of the universal form for a whole class of gauge theories. However, the case of Abelian fields differs from that of nonAbelian ones. In the electromagnetic theories the gauge boson and scalar actions are quadratic in fields.  Therefore, the $n-$point functions generated by these actions vanish for $n>2$. Only the fermion action generates the non-trivial three-point and higher functions. The action (\ref{action-Psi-HL}) is, in particular, responsible for a modification of the electromagnetic vertex. In the nonAbelian theories, both the gauge boson and fermion actions generate  the non-trivial three-point and higher functions. Therefore, the gluon-fermion, three-gluon, and four-gluon couplings are all modified.

%%%%%%%%%%%%%%%%%%%%%%%%%%%%%%%%%%%%%%%%%%%%%
\section{Discussion}
\label{sec-discussion}
%%%%%%%%%%%%%%%%%%%%%%%%%%%%%%%%%%%%%%%%%%%%%

We have shown that the hard-loop self-energies of gauge, fermion, and scalar fields are of the universal structures and so are the effective actions of QED, scalar QED, $\mathcal{N}=1$ super QED, Yang-Mills, QCD, and $\mathcal{N}=4$ super Yang-Mills. One asks why the universality occurs physically. Taking into account a diversity of the theories - various field content and microscopic interactions - the uniqueness of the hard-loop effective action is rather surprising. 

The universality of hard-loop action means that neither effects of quantum statistics of plasma constituents are observable nor the differences in elementary interactions which govern the dynamics of the two systems. The hard-loop approximation requires that the momentum at which a plasma is probed, that is the wavevector $k$, is much smaller than the typical momentum of a plasma constituent $p$. Therefore, the length scale, at which the plasma is probed, $1/k$, is much greater than the characteristic de Broglie wavelength of plasma particle, $1/p$. The hard-loop approximation thus corresponds to the classical limit where fermions and bosons of the same masses and charges are not distinguishable. The fact that the differences in elementary interactions are not seen results from the very nature of gauge theories - the gauge symmetry fully controls the interaction. And the hard-loop effective actions obey the gauge symmetry. 

The universality of hard-loop actions has far-reaching physical consequences: the characteristics of all plasma systems under consideration, which occur at the soft scale, are qualitatively the same. In particular, spectra of collective excitations of gauge, fermion, and scalar fields are the same. Therefore, if the electromagnetic plasma with a given momentum distribution is, say, unstable, the quark-gluon plasma with this momentum distribution is unstable as well. We conclude that in spite of all differences, the plasma systems under consideration are very similar to each other at the soft scale. Below the magnetic sale these systems can behave very differently. 

\vspace{4mm}

This work was partially supported by the Polish National Science Centre under Grant No. 2011/03/B/ST2/00110.

%--------------------------------------------------------------------------------------------------------------------------


\begin{thebibliography}{99}
%--------------------------------------------------------------------------------------------------------------------------

%\cite{Thoma:1995ju}
\bibitem{Thoma:1995ju}
M.~H.~Thoma,
%``Applications of high temperature field theory to heavy ion collisions,''
in {\it Quark-Gluon Plasma 2,} edited by R.C.~Hwa
(World Scientific, Singapore, 1995).
%arXiv:hep-ph/9503400.
%%CITATION = HEP-PH/9503400;%%

%\cite{Blaizot:2001nr}
\bibitem{Blaizot:2001nr}
J.~P.~Blaizot and E.~Iancu,
%``The quark-gluon plasma: Collective dynamics and hard thermal loops,''
Phys.\ Rept.\  {\bf 359}, 355 (2002).
%[arXiv:hep-ph/0101103].
%%CITATION = PRPLC,359,355;%%

%\cite{Litim:2001db}
\bibitem{Litim:2001db} 
D.~F.~Litim and C.~Manuel,
%``Semiclassical transport theory for nonAbelian plasmas,''
Phys.\ Rept.\  {\bf 364}, 451 (2002).
%[hep-ph/0110104].
%%CITATION = HEP-PH/0110104;%%
  
%\cite{Kraemmer:2003gd}
\bibitem{Kraemmer:2003gd}
U.~Kraemmer and A.~Rebhan,
%``Advances in perturbative thermal field theory,''
Rept.\ Prog.\ Phys.\  {\bf 67}, 351 (2004).
%[arXiv:hep-ph/0310337].
%%CITATION = RPPHA,67,351;%%

%\cite{Braaten:1989mz}
\bibitem{Braaten:1989mz} 
E.~Braaten and R.~D.~Pisarski,
%``Soft Amplitudes in Hot Gauge Theories: A General Analysis,''
Nucl.\ Phys.\ B {\bf 337}, 569 (1990).
%%CITATION = NUPHA,B337,569;%%

%\cite{Braaten:1990az}
\bibitem{Braaten:1990az} 
E.~Braaten and R.~D.~Pisarski,
%``Deducing Hard Thermal Loops From Ward Identities,''
Nucl.\ Phys.\ B {\bf 339}, 310 (1990).
%%CITATION = NUPHA,B339,310;%%

%\cite{Blaizot:1993be}
\bibitem{Blaizot:1993be}
J.~P.~Blaizot and E.~Iancu,
%``Soft collective excitations in hot gauge theories,''
Nucl.\ Phys.\ B {\bf 417}, 608 (1994).
%[arXiv:hep-ph/9306294].
%%CITATION = HEP-PH 9306294;%%

%\cite{Kelly:1994dh}
\bibitem{Kelly:1994dh}
P.~F.~Kelly, Q.~Liu, C.~Lucchesi and C.~Manuel,
%``Classical transport theory and hard thermal loops in the quark - gluon plasma,''
Phys.\ Rev.\ D {\bf 50}, 4209 (1994).
%[arXiv:hep-ph/9406285].
%%CITATION = HEP-PH 9406285;%%

%\cite{Pisarski:1997cp}
\bibitem{Pisarski:1997cp}
R.~D.~Pisarski,
%``Nonabelian Debye screening, tsunami waves, and worldline fermions,''
arXiv:hep-ph/9710370.
%%CITATION = HEP-PH/9710370;%%

%\cite{Mrowczynski:2000ed}
\bibitem{Mrowczynski:2000ed}
St.~Mr\'owczynski and M.~H.~Thoma,
%``Hard loop approach to anisotropic systems,''
Phys.\ Rev.\  D {\bf 62}, 036011 (2000).
%[arXiv:hep-ph/0001164].
%%CITATION = PHRVA,D62,036011;%%

%\cite{Taylor:1990ia}
\bibitem{Taylor:1990ia}
J.~C.~Taylor and S.~M.~H.~Wong,
%``THE EFFECTIVE ACTION OF HARD THERMAL LOOPS IN QCD,''
Nucl.\ Phys.\  B {\bf 346}, 115 (1990).
%%CITATION = NUPHA,B346,115;%%

%\cite{Frenkel:1991ts}
\bibitem{Frenkel:1991ts}
J.~Frenkel and J.~C.~Taylor,
%``Hard thermal QCD, forward scattering and effective actions,''
Nucl.\ Phys.\  B {\bf 374}, 156 (1992).
 %%CITATION = NUPHA,B374,156;%%

%\cite{Braaten:1991gm}
\bibitem{Braaten:1991gm}
E.~Braaten and R.~D.~Pisarski,
%``Simple effective Lagrangian for hard thermal loops,''
Phys.\ Rev.\  D {\bf 45}, 1827 (1992).
%%CITATION = PHRVA,D45,1827;%%

%\cite{Mrowczynski:2004kv}
\bibitem{Mrowczynski:2004kv}
St.~Mr\'owczy\'nski, A.~Rebhan and M.~Strickland,
%``Hard-loop effective action for anisotropic plasmas,''
Phys.\ Rev.\  D {\bf 70}, 025004 (2004).
%[arXiv:hep-ph/0403256].
%%CITATION = PHRVA,D70,025004;%%

%\cite{Weldon:1982aq}
\bibitem{Weldon:1982aq} 
H.~A.~Weldon,
%``Covariant Calculations at Finite Temperature: The Relativistic Plasma,''
Phys.\ Rev.\ D {\bf 26}, 1394 (1982).
%%CITATION = PHRVA,D26,1394;%%

%\cite{Mrowczynski:2007hb}
\bibitem{Mrowczynski:2007hb}
St.~Mr\'owczy\'nski and M.~H.~Thoma,
%``What do electromagnetic plasmas tell us about quark-gluon plasma?,''
Ann.\ Rev.\ Nucl.\ Part.\ Sci.\  {\bf 57}, 61 (2007).
%[arXiv:nucl-th/0701002].
%%CITATION = ARNUA,57,61;%%

%\cite{Czajka:2014gaa}
\bibitem{Czajka:2014gaa} 
A.~Czajka and St.~Mr\'owczy\'nski,
%``Universality of the hard-loop action,''
Phys.\ Rev.\ D {\bf 91}, 025013 (2015).
%[arXiv:1410.5117 [hep-ph]].
%%CITATION = ARXIV:1410.5117;%%

%\cite{Czajka:2010zh}
\bibitem{Czajka:2010zh} 
A.~Czajka and St.~Mr\'owczy\'nski,
%``Collective Excitations of Supersymmetric Plasma,''
Phys.\ Rev.\ D {\bf 83}, 045001 (2011).
%[arXiv:1011.6028 [hep-th]].
%%CITATION = ARXIV:1011.6028;%%

%\cite{Czajka:2011zn}
\bibitem{Czajka:2011zn} 
A.~Czajka and St.~Mr\'owczy\'nski,
%``Collisional Processes in Supersymmetric Plasma,''
Phys.\ Rev.\ D {\bf 84}, 105020 (2011).
%[arXiv:1108.2388 [hep-th]].
%%CITATION = ARXIV:1108.2388;%%

%\cite{Czajka:2012gq}
\bibitem{Czajka:2012gq} 
A.~Czajka and St.~Mr\'owczy\'nski,
%``N=4 Super Yang-Mills Plasma,''
Phys.\ Rev.\ D {\bf 86}, 025017 (2012).
%[arXiv:1203.1856 [hep-th]].
%%CITATION = ARXIV:1203.1856;%%

%\cite{Mrowczynski:1992hq}
\bibitem{Mrowczynski:1992hq}
St.~Mr\'owczy\'nski and U.~W.~Heinz,
%``Towards a relativistic transport theory of nuclear matter,''
Annals Phys.\  {\bf 229}, 1 (1994).
%%CITATION = APNYA,229,1;%%

%\cite{Lebedev:1989ev}
\bibitem{Lebedev:1989ev} 
V.~V.~Lebedev and A.~V.~Smilga,
%``Spectrum of Quark - Gluon Plasma,''
Annals Phys.\  {\bf 202}, 229 (1990).
%%CITATION = APNYA,202,229;%%

\end{thebibliography}
\end{document}